\documentclass[
journal=jacsat, % for undefined journal
manuscript=article]{achemso}

\usepackage[version=3]{mhchem} % Formula subscripts using \ce{}
\usepackage{siunitx}

\author{Johanna Salvant}
\altaffiliation{Current address: C2RMF, Palais du Louvre, Porte des Lions, 14, quai François Mitterrand, Paris}
\author{Marc Walton}
\email{marc.walton@northwestern.edu}
\affiliation[Northwestern University]
{Center for Scientific Studies in the Arts, Northwestern University, Evanston IL, USA}
\author{Dale Kronkright}
\affiliation[Georgia O'Keeffe Museum, Santa Fe, NM, USA]{Georgia O'Keeffe Museum, Santa Fe, NM, USA}
\author{Chia-Kai Yeh}
\author{Fengqiang Li}
\author{Oliver Cossairt}
\author{Aggelos K. Katsaggelos}
\affiliation[Northwestern University]
{Electrical Engineering and Computer Science, Northwestern University, Evanston, IL, USA}

\title[\texttt{achemso} demonstration]
{Photometric Stereo by UV-Induced Fluorescence to Detect Protrusions on Georgia O'Keeffe's Paintings}

\begin{document}

\begin{abstract}
\setlength{\parindent}{4em}
\setlength{\parskip}{1em}
\renewcommand{\baselinestretch}{1}
A significant number of oil paintings produced by Georgia O'Keeffe (1887-1986) show surface protrusions of varying width, up to several hundreds of microns. These protrusions are similar to those described in the art conservation literature as metallic soaps. Since the presence of these protrusions raises questions about the state of conservation and long-term prospects for deterioration of these artworks, a 3D-imaging technique, photometric stereo using ultraviolet illumination, was developed for the long-term monitoring of the surface-shape of the protrusions and the surrounding paint. Because the UV fluorescence response of painting materials is isotropic, errors typically caused by non-Lambertian (anisotropic) specularities when using visible reflected light can be avoided providing a more accurate estimation of shape. As an added benefit, fluorescence provides additional contrast information contributing to materials characterization. The developed methodology aims to detect, characterize, and quantify the distribution of micro-protrusions and their development over the surface of entire artworks. Combined with a set of analytical in-situ techniques, and computational tools, this approach constitutes a novel methodology to investigate the selective distribution of protrusions in correlation with the composition of painting materials at the macro-scale. While focused on O'Keeffe's paintings as a case study, we expect the proposed approach to have broader significance by providing a non-invasive protocol to the conservation community to probe topological changes for any relatively flat painted surface of an artwork, and more specifically to monitor the dynamic formation of protrusions, in relation to paint composition and modifications of environmental conditions, loans, exhibitions and storage over the long-term.
\end{abstract}

\section{Introduction}

Metal soap-protrusions are a form of deterioration that affects scores of oil-based paintings made from antiquity until the present day (Noble et al. 2002; Higgitt et al. 2003; Noble et al. 2005; Keune 2005; Jones et al. 2007; Ferreira et al. 2011). Understanding how metal soap protrusions form and develop over time constitutes a major challenge to painting conservation, as this widespread phenomenon is known to alter the visual appearance of the painted surface (Noble et al. 2005; Noble and Boon 2007; Shimadzu and Van den Berg 2006; Centeno and Mahon 2009) and to compromise its chemical and mechanical stability (Rogala et al. 2010; Maines et al. 2011). A growing body of literature has emerged on the investigation of metal soap protrusions in paintings over the past twenty years. While most of this research has focused on investigating metal soap protrusions at the micro- and molecular scale using a variety of techniques requiring microsamples (Heeren et al. 1999; Noble et al. 2002; van der Weerd 2002; Higgitt et al. 2003; Plater et al. 2003; Keune 2005; Osmond et al. 2005; Keune et al. 2005; Keune and Boon 2007; Cotte et al. 2007; Spring et al. 2008; Ferreira et al. 2011; Osmond et al. 2012; Osmond 2014; Ferreira et al. 2015), in this study an easy-to-implement and non-invasive methodological approach is described, using artworks by Georgia O'Keeffe as models, to document the macro distribution of protrusions. 

Based on a survey initiated by the Georgia O'Keeffe Museum in 2009, a subset of O'Keeffe's paintings produced between 1920 and 1950 were identified as having disfiguring micro-protrusions scattered across their surfaces. These protrusions exhibit a strong UV-induced fluorescence response, range in size from 10 to greater than \SI{200}{\micro\metre}, and occasionally appear erupted with a “caldera-like” shape consistent with soap aggregates found in other modern and early modern paintings (Osmond et al. 2005; O'Donoghue et al. 2006; Faubel et al. 2011; Ferreira et al. 2011; Duffy et al. 2014; Helwig et al. 2014). The protrusion formation process must have started at an early stage in these artworks’ history: in a 1947 correspondence between O'Keeffe and conservator Caroline Keck (Keck 1947), the artist mentioned that she noticed opaque, granular textures and pinpoint losses appearing in several of her oil paintings created between 1928-1936, some of which today exhibits particularly large protrusions. The identification of these protrusions in the 2009 survey led to a number of questions about when they first developed on these artworks and whether they were still actively changing. 
The presence and potential ongoing growth of protrusions on paintings raises significant concerns for their long-term preservation. Characterizing the protrusions, their distribution over the surface of the artworks and monitoring their development in time has now become a conservation priority not only for O'Keeffe's oeuvre, but for all paintings affected by protrusions. Likewise, investigating the mechanisms and factors promoting the protrusion development in artworks has become essential, especially since the collections are often used in travelling exhibitions and are subjected to environmental fluctuations that may contribute to the ongoing development of protrusions.

\begin{figure}[h!]
  \includegraphics[width=\textwidth,height=\textheight,keepaspectratio]{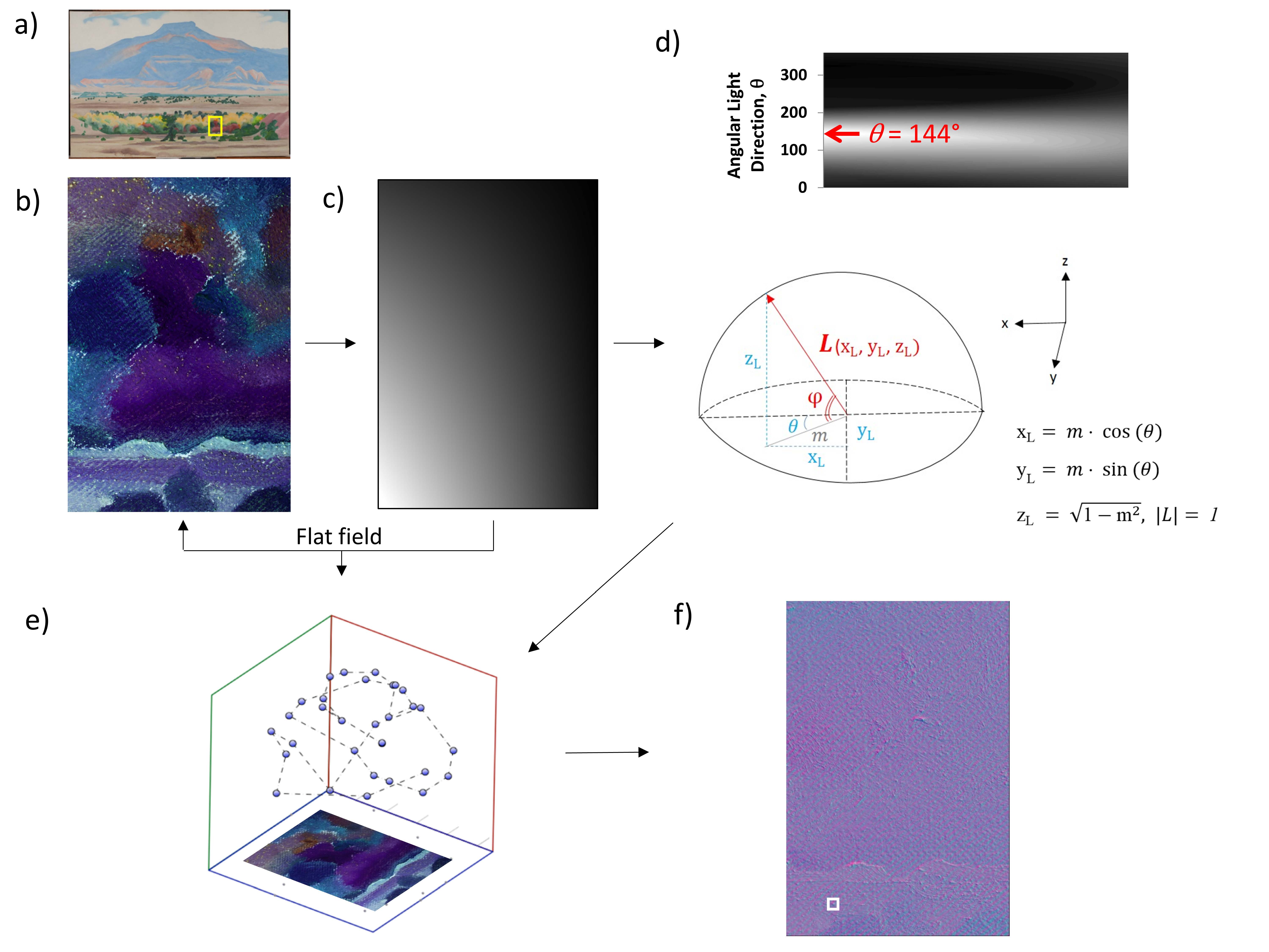}
  \caption{Pre-processing pipeline of photometric stereo data: a) Georgia O’Keeffe, Pedernal, 1941/42, oil on canvas, CR1029, Georgia O’Keeffe Museum, Santa Fe, NM. Region selected as representative example for investigation is shown by yellow rectangle; a single image illuminated with a directional light b) was fit with an orthogonal Legendre polynomial to calculate a map of light drop-off c) which shows the light originating from the the direction of the lower left corner of the image. To achieve a precise measure of the lighting direction, polynomial-fit image was reshaped into polar coordinates d) to determine the angle $θi$ of the illumination vector $Li$. The X,Y, Z position of each illumination direction e) was determined using an optimization procedure described in the text. Photometric stereo was calculated using linear least squares to produce a normal vector map f). The box demarcates the detail region shown in figure 2.}
\end{figure}

In this work, in order to document the dynamic development of protrusions over time, a novel photometric stereo (PS) protocol was developed to image the 3D-texture of the painted surface where fine features of approximately \SI{10}{\micro\metre} wide can be resolved. We achieve these figures of merit by exploiting the fluorescence properties of pigments and binding media when excited by ultraviolet wavelengths. This isotropic response (Rost 1995) is ideally suited to using PS, which assumes Lambertian (diffuse) reflectance from a surface (Woodham 1978). As a second step, we also describe two data processing methods for detection of the protrusions using the PS data as an input: a manual count is compared to a data driven approach. These counts of protrusions in different regions as well as the measurement of their size distribution, are then correlated with paint composition.

\subsection{Experimental Capture and Processing Procedures}

The Georgia O'Keeffe oil painting Pedernal (figure 1a) has been chosen as a case study to illustrate the processing and analysis of PS data. Imaging was undertaken on a region of the painting, demarcated in figure 1a, that exhibits a high number of protrusions and is representative of the overall palette. The region size of approximately 7.6 x 5.0cm (height x width) was chosen to achieve high spatial resolutions of approximately \SI{10}{\micro\metre} using our camera/lens configuration. Specifically, in the conditions used, spatial sampling of approximately \SI{15}{\micro\metre} per pixel over a field of view of 4 cm was achieved. It should also be noted that exposure to these UV wavelengths was kept at a minimum, similar to that needed for a UV photograph of a painted suface, to ensure the safety of the painting. The setup utilizes a 365nm UV-lamp (fitted with a Sylvania F15T8/BLB bulb) as the illuminant and a Canon EOS 5D Mark III camera equipped with a 50mm prime lens.

\subsection{Photometric Stereo data acquisition and processing}
\subsubsection{Capture}
At its core, our method utilizes a common camera configuration described previously for PS (Woodham 1978) and reflectance transformation imaging (RTI) measurements; the camera is fixed orthogonal to the artwork and a series of photographs are taken with different angles of illumination (Malzbender et al. 2001; Padfield et al. 2005; Mudge et al. 2005; Mudge et al. 2006; Fattal et al. 2007; Mudge et al. 2008; Earl et al. 2010a; Earl et al. 2010b; Earl et al. 2011; Malesevic et al. 2013; Duffy et al. 2013). Following the well known photometric stereo equation, albedo (color) and normal maps (shape) of the object surface are recovered: 
\begin{equation}
I =\mathcal{N}\cdot L  ,\  \mathcal{N}̃=kN,
\end{equation}
	
In this parameterization, the albedo (color) $k$ is absorbed into the surface normal so that $|\mathcal{N}| = k$. Also the image intensity is given by $I(x,y)$ under illumination vector $L$.
Shape recovery in PS requires that the surfaces being measured scatter light diffusely, often called Lambertian scattering. However, few materials exhibit ideal diffuse behavior but instead reflect light specularly. Painted surfaces often exhibit both diffuse as well as specular properties. These non-ideal surfaces are difficult to model accurately using PS alone (Ma et al. 2007), so recent studies have highlighted the advantages of using fluorescence-based PS for 3D-shape reconstruction compared to previous methods using reflected light (Treibitz et al. 2012; Sato et al. 2012). Stokes fluorescence emissions induced by a directional short wavelength light are isotropically radiated at longer wavelengths with ideal Lambertian characteristics where the emission intensity varies according to the cosine of the lighting angle (Kratohvil et al. 1978; Gordon et al. 1993; Glassner 1995). Therefore the use of fluorescence emission as the source of light for PS of painted surfaces can enable more accurate, precise, and repeatable shape measurements than the use of reflected light. 

\subsubsection{Pre-processing}
\subsubsection{Calculating the illumination vector}
In a typical “free-form” PS or reflectance transformation imaging capture set-up, a reflective ball is used to record the position of a far-light source $L$ (Mudge et al. 2008). Here we have eliminated the need for using the reflective ball through an algorithm designed for near-lighting conditions (Huang et al. 2015; Cossairt et al. 2015). One of the principal benefits of removing the mirror ball and other calibration hardware from the scene is that the entire field of view may be dedicated to capturing the desired image. Also, this method where the light position is estimated directly from the object itself exploiting the inverse square law of light drop-off may be more accurate than using a mirror ball (Huang et al. 2015). This is especially true if the material being measured is a true Lambertian scatterer as is the case of fluorescence measurements (Treibitz et al. 2012). Lastly, the weak intensity of most fluorescent sources necessitates near light conditions in order to provide enough photon flux to produce an acceptable signal to noise characteristic in the captured images. In our example, the light was positioned about 500cm away from the surface.

In comparison to Huang et al. (2015), in this study a different image processing pipeline is used which is more user-friendly for conservators, museum imaging departments, and conservation scientists interested in adopting the framework described. All of these steps were performed with either open-source and off-the-shelf software, such as ImageJ (Schneider et al. 2012; Schindelin et al. 2015). The code used in this project is available from an online repository (https://github.com/NU-ACCESS/ImageJ-Photometric-Stereo-Tools) as Python scripts that can be readily downloaded and installed into ImageJ. 

Light position estimation was achieved in a three-step process, as illustrated in figure 1b-f. First, each lighting direction image (figure 1b) was fit with an orthogonal polynomial in ImageJ, as exemplified in figure 1c, that is normalized to the mean gray value. While, polynomial fitting of an image is typically used to compensate for uneven illumination, here we use the fit-image (figure 1c) to estimate illumination direction. The fit-image can be re-shaped into polar coordinates by radially slicing the image (figure 1d), from which the angle of illumination θ can be directly extracted from the point of maximum intensity. Second, the polynomial image is used to normalize or ‘flat-field’ each lighting direction image to correct for the unevenness of illumination induced by the near light. Third, we borrow from the near-light model (Huang et al. 2015), in which the Cartesian coordinates ($x_L, y_L, z_L$) of the vector of illumination $L$ are calculated from the direction and magnitude of the light drop-off. It is assumed that all light source positions are equidistant to the center of the region of investigation (figure 1d), meaning $|L|=1$.

As illustrated in figure 1e, each lighting direction $L$ is determined by iteratively adjusting a parameter $m$ while updating the surface normal $\mathcal{N}$. This is performed until the difference between $L(m)\cdot\mathcal{N}$ and $I$ becomes very small as per this cost function: 

\begin{equation}
\operatorname*{argmin}_{\mathcal{N},m}|L(m)\cdot\mathcal{N}-I|
\end{equation}

\subsubsection{Calculating surface normal, depth, and albedo maps}
Based on the estimated vectors of illumination $L$, the PS equation (1) is solved using least squares to recover the albedo and surface normal maps, the latter shown in figure 1f. Examples of x and y gradients, depth and albedo images are illustrated in figures 2a-d as high resolution details from Pedernal. The depth map (figure 2c) is calculated by integrating the x and y gradients using standard algorithms such as described by Frankot and Chellappa (1988). Clearly apparent in the center of the image is a large protrusion (approx. \SI{200}{\micro\metre} across) with other smaller protrusions (30-\SI{150}{\micro\metre}) peppered across the surface that have whitish to yellowish fluorescence (Fig. 2d). Figure 2e illustrates the 3D-visualization of the depth map that enables characterization of the morphological features of the protrusions.

\begin{figure}[h!]
  \includegraphics[width=\textwidth,height=\textheight,keepaspectratio]{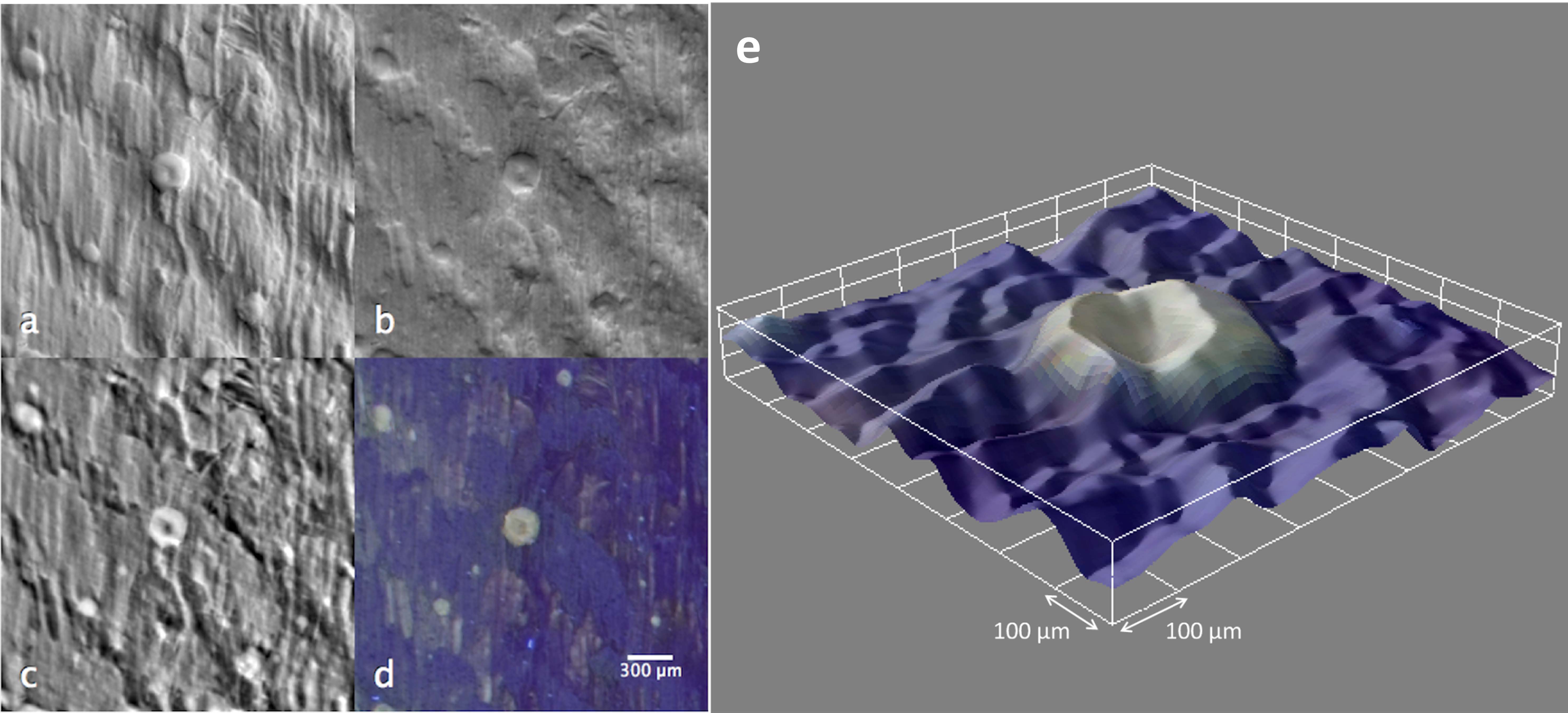}
  \caption{After the flat-fielding procedure describe by figure 1, the complete set of racking light images were used to calculate normal vector, depth, and albedo images. Details (a-d) of an area of 300 x 300 pixels (4 x 4 mm) show the X-gradient (a) and Y-gradient (b) of the surface normal vector N, the depth map (c), and albedo (d) providing information on the diffuse color generated by the UV induced fluorescence of the surface with underling contrast from shape removed. The 3D rendering (e) shows a ‘caldera’ shape of the central protrusion.}
\end{figure}

\subsection{Protrusion detection}
The use of UV-induced fluorescence presents two benefits in addition to the improved accuracy of the PS measurements. First, the calculated albedo  is a perfectly diffuse and flat record of fluorescence which aids in the accurate detection and segmentation of the protrusions. Second, information on the UV-induced fluorescence response of the protrusions and of surrounding material is simultaneously collected in the visible range, which can contribute to materials differentiation and characterization. 
Using both the albedo and shape information, two data analysis methods were tested and compared to investigate the occurrence and distribution of protrusions in different colored areas. The first method, which is manual, uses ImageJ to detect protrusions on the albedo image with local color thresholding. The albedo image allows for a convenient visualization of the protrusions, as shown in figure 2d, due to their strong fluorescence response. This method provides quantitative information on the protrusions by detecting all visible protrusions within the investigated region. The second method is a fully automated detection scheme using an experimental algorithm implemented with Matlab and OpenCV. This algorithm has been designed for un-aided detection of the protrusions, based on the combination of characteristic colors and surface features. 

\subsubsection{Manual Protrusion Detection}
Visible image of the investigated region is shown in figure 3a. For each observable protrusion in the albedo image, contiguous pixels of similar RGB values within an optimized adjusted range are hand-selected to fit the outline of the protrusion area (via the “wand tool”). All selected pixels are used to create a map of the protrusion distributions illustrated as a red overlay on the image of the fluorescence response (figure 3b). 
The visible image (figure 3a) was manually segmented to isolate areas that appeared to have similar colors based on the visible and albedo images, based on the observed colors (figure 3c). Each of these areas was masked (figure 3d) to count the total number of protrusions per color. This, together with their location, size, and area were all recorded (figure 3e). The process was repeated for each segmented color area. 
Despite being time-consuming, the manual method is advantageous since it is very easy to implement and accurate for detection of protrusions larger than \SI{10}{\micro\metre}, considering the resolution of the PS data used in the current example. 

\begin{figure}[h!]
  \includegraphics[width=\textwidth,height=\textheight,keepaspectratio]{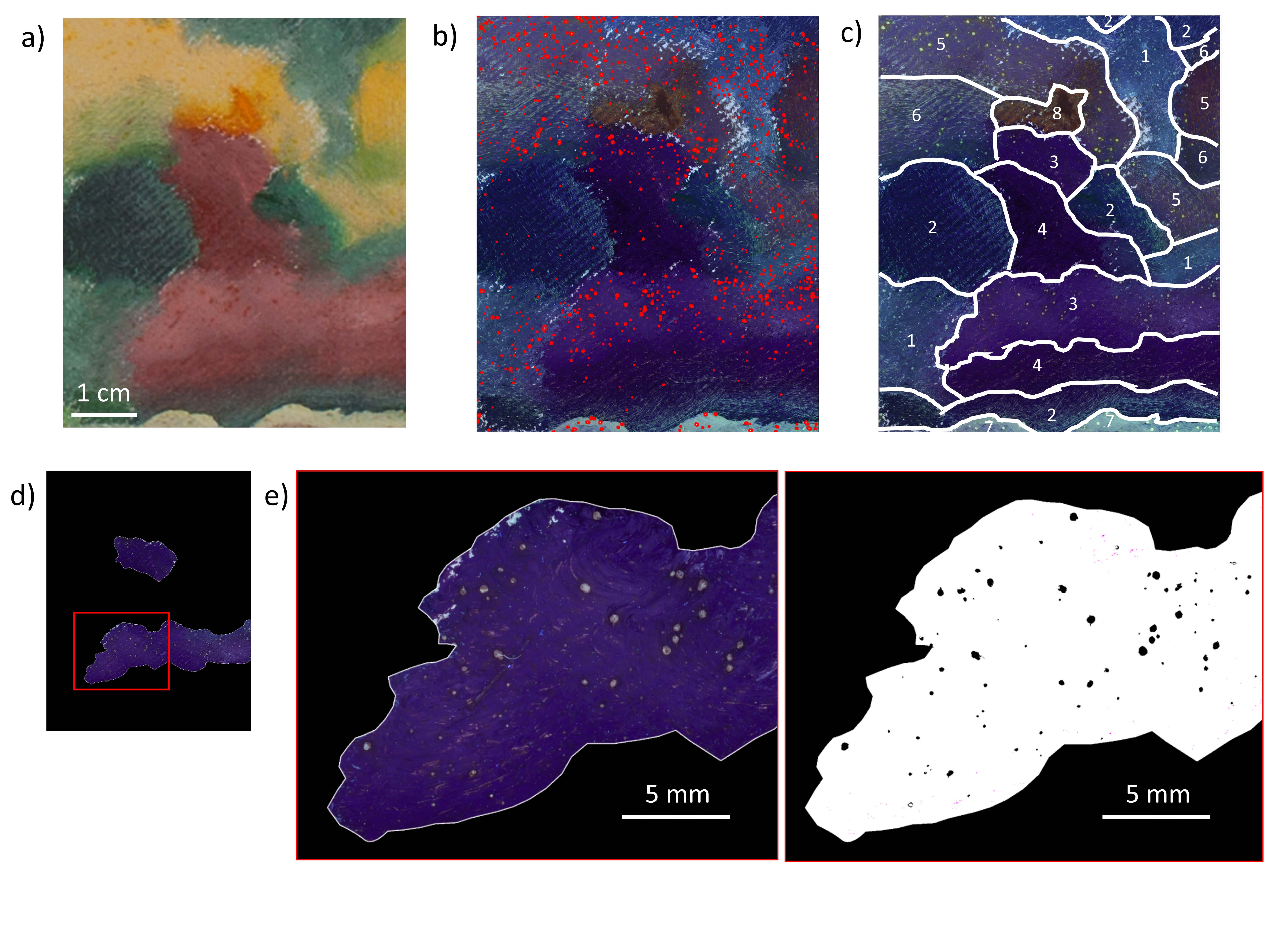}
  \caption{a) Visible image of the area (7.5 x 5.0 cm) from Pedernal to test methods for data analysis; b) UV-albedo image of the same region with red overlay mask of the protrusions manually selected using ImageJ; c) UV-albedo image with schematic indication of the eight different colored area for segmentation; d) region 4 masked from the UV-albedo image; e) steps for protrusions analysis in ImageJ: detail of light red colored mask (left); results for that detail from multiplying the mask by the overlay of detected protrusions, showing the protrusions in black in the red light area (middle); image resulting from particles analysis where the counted protrusions are outlined in black and counted in red.}
\end{figure}

\subsubsection{Automated Protrusion Detection}
A custom algorithm was applied to test automated quantification of the protrusions in the images acquired by PS. As for the manual method, each color area was manually segmented and masked (figures 3c-d). For the automated detection algorithm, the masks were then applied to both the albedo and normal images. Color thresholding and template matching on the albedo image was used to generate a binary mask, where a value of 1 indicates the presence of a protrusion. The normal image was integrated using the Frankot-Chellappa algorithm to produce a depth/height map of the surface of the painting. The OpenCV blob detection routine was then used to locate protrusions in the depth image, generating a second binary mask. A composite mask is then generated from the union of the two binary masks. The same procedure for analyzing protrusion statistics within each color region is then applied. The combination of information from the color albedo and surface shape images enables better rejection of false positive protrusion detection, such as those that may be caused by brushwork texture or areas where the primer is apparent.
For both methods, the number of protrusions/$cm^2$ was calculated as a function of protrusion width for each color area (Fig. 4a).

\subsection{X-ray fluorescence (XRF) and Portable Fourier transformed infrared spectroscopy (FTIR)}
XRF imaging was carried out on the same region captured during the PS analysis (figure 1a). Using an ELIO X-ray fluorescence imaging spectrometer (XGLab), equipped with a Rh tube and 1mm spot size. Rastering was executed with a \SI{250}{\micro\metre} step size and 1.6s acquisition time for each point. The instrument was operated at 50kV and 40μA. The open access PyMCA software (Solé et al. 2007) was used to fit the XRF map and obtain the distribution of each element.

FTIR spectra were collected for each painted color area found in the region of interest using a Bruker Alpha small footprint portable FTIR spectrometer, in reflection mode sampling and spectral range 6000-$400cm^{-1}$, a measurement spot of 6mm in diameter, and working distance of ~15mm. 256 scans were acquired with a $4cm^{-1}$ resolution.

\section{Results and Discussion}
\subsection{Detection of protrusions}
\subsubsection{“Ground truth” manual method}
The red overlay on the fluorescence image (figure 3b) shows the distribution of protrusions across the painted surface. Protrusions appear most abundant in areas corresponding to light tonalities (e.g., light red, light green, beige) whereas they are almost absent in other darker tonalities (e.g., dark green, dark red). Frequency histograms (figure 4a) show the number of protrusions/$cm^2$ as a function of their width. From these data, the beige, light green, light red, yellow and yellow green areas have a high number of protrusions (34 to 57 protrusions/$cm^2$). In contrast, the number of protrusions in the dark green, dark red and orange areas is significantly lower, never exceeding 10 protrusions/$cm^2$. As will be discussed in more detail below, these data show that lighter and darker shades of the same colors (as in dark vs. light green- likely made from the same chromium-based pigment as determined by XRF), clearly show dissimilar protrusion counts. 
The profiles of size distributions are overall similar for the different colored areas, with a dominant proportion of protrusions in the 100-\SI{200}{\micro\metre} range. Only the dark red and yellow green areas exhibit a slightly higher proportion of small protrusions less than \SI{100}{\micro\metre} wide. It can also be noted that the areas with abundant protrusions skewed to distributions with larger protrusions- with widths often exceeding \SI{400}{\micro\metre}. It should be noted that for this proof-of-concept work, relatively large bins of \SI{100}{\micro\metre} have been chosen to illustrate the capabilities of the method and reduce the number of false positives; however, in the future, smaller bins. Especially in the 0-\SI{300}{\micro\metre} range should be considered to better characterize the variety of sizes of the protrusions.

\begin{figure}[h!]
  \includegraphics[width=\textwidth,height=\textheight,keepaspectratio]{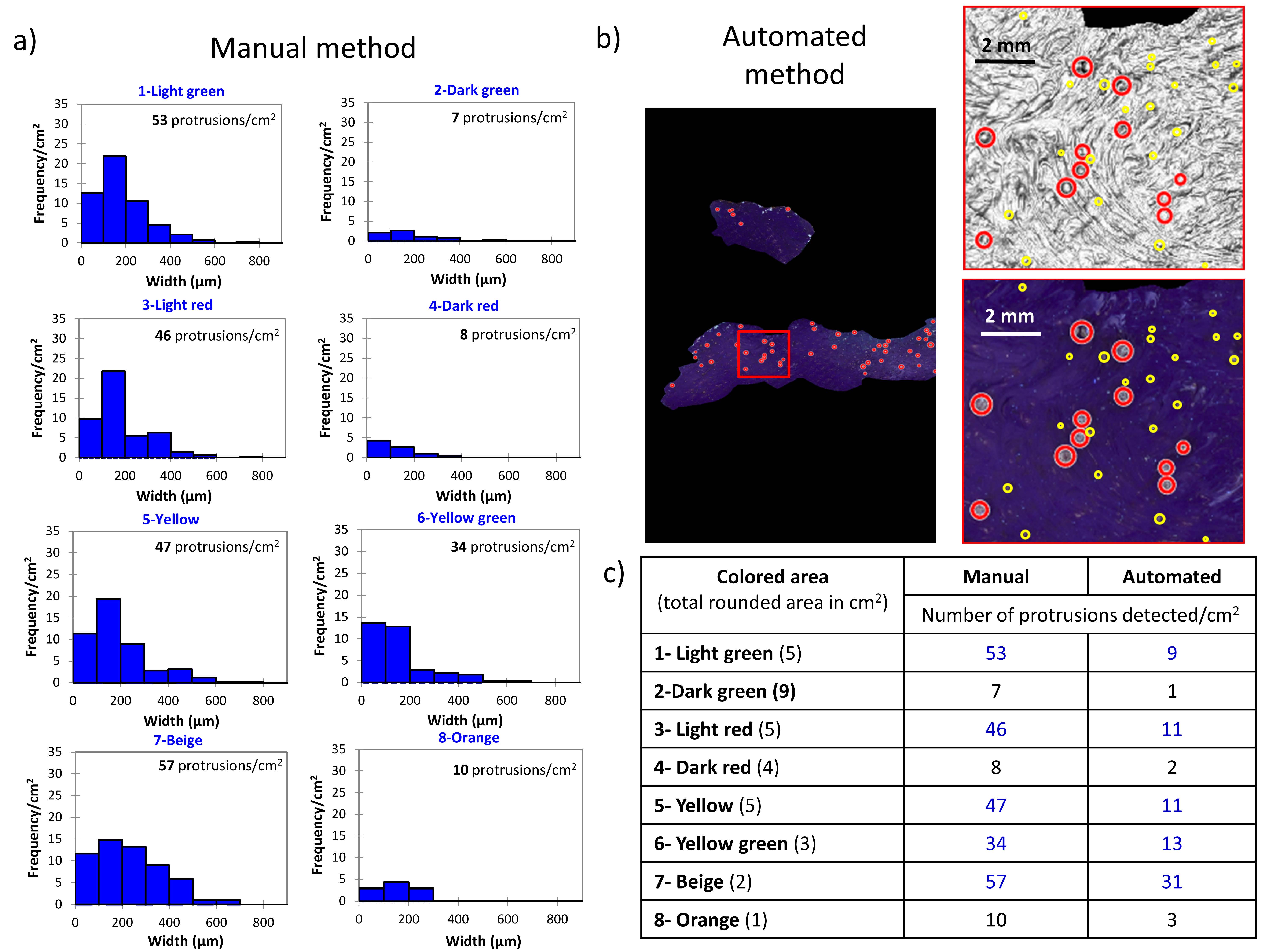}
  \caption{Results of protrusions analysis and comparison of  methods: a) Results of ground true semi-automatic method of analysis. Histogram of protrusion occurrence and range of size for the eight colored areas; b) results from the computer automatic method for the yellow region: mask (top); details of results of protrusions detection (circle in red) from the depth z map calculated from normal map (bottom left) and from the albedo map (bottom right); c) table comparing the occurrence of protrusions detected (number/$cm^2$) for each colored area in the investigated region by the two methods.}
\end{figure}

\subsubsection{Evaluation of the automated method}
Figure 4b shows the detected protrusions, circled in red, for the light red area using the automated algorithm overlaid onto the depth map and albedo images. For comparison, the manually detected protrusions missed by the automated method are circled in yellow. The detection of relatively big protrusions is fairly robust while detection of small protrusions (approx. \SI{200}{\micro\metre}) is not accurate. The challenge in detecting small protrusions lies in the fact that false positives are generated from the surface texture produced by the brush marks and from the white priming layer, contrasting in color with paint materials in areas where it remains visible. Nevertheless, as shown in figure 4c, the automated method is still able to correctly attribute the higher frequency of protrusions to lightly colored areas, as does the manual method but with significantly reduced processing time. Further development of this preliminary version of the automated algorithm is however necessary to optimize detection accuracy so that it can be reliably applied to monitoring the variation of protrusion statistics over time. 

\subsection{Correlating protrusions with paint composition at the macroscale}
Non-invasive FTIR and XRF analysis were performed on Pedernal as a case study to investigate how the distribution of protrusions correlates with paint materials. Many of O’Keeffe’s artworks that exhibit protrusions have been painted on a specific type of commercial pre-primed canvas. Based on this observation, it was originally speculated by the museum conservators that the priming layer could have played a role in the protrusions formation and was thus a focus of our investigation. As explained below, our analysis determined that, on the contrary, the distribution of protrusions correlate with the composition and color of paint layers, and thus the degradation phenomena observed at the surface cannot be attributed to the priming. 

FTIR analysis of the bare commercial primer (figure 5a) indicates that the primer is composed of lead white (mixture of hydrocerussite (\ce{Pb3(CO3)2(OH)2}) and cerussite (\ce{PbCO3})), barium sulfate (\ce{BaSO4}) and calcium carbonate (\ce{CaCO3}), based on the presence of the strong derivative band of the asymmetric \ce{CO3^{2-}} stretching at ca. 1465$cm^{-1}$ and the characteristic stretching and bending vibrational modes of hydrocerussite (3550, 1045, 680$cm^{-1}$) and combination and overtone modes of cerussite ($2413cm^{-1}$) and of calcium carbonate (4260, 2511, $1794cm^{-1}$) (Miliani et al. 2012). It should be noted that SEM/EDX analysis of a cross section from Pedernal indicated that the priming is primarily composed of chalk with barium sulfate, and only minor quantities of lead compounds, which are, on the contrary, dominant in the paint layers. The sharp CH stretching at 2922 and $2846cm^{-1}$ combined with the doublets in the 1540-1520 and 730-720$cm^{-1}$ regions, as shown in the details of figure 5a, point out towards the presence of metal carboxylates (Robinet and Corbeil 2003). Only few visible protrusions are observed on the unpainted commercial primer and when they are detected they tend to be small (\SI{150}{\micro\metre}). 

\begin{figure}[h!]
  \includegraphics[width=\textwidth,height=\textheight,keepaspectratio]{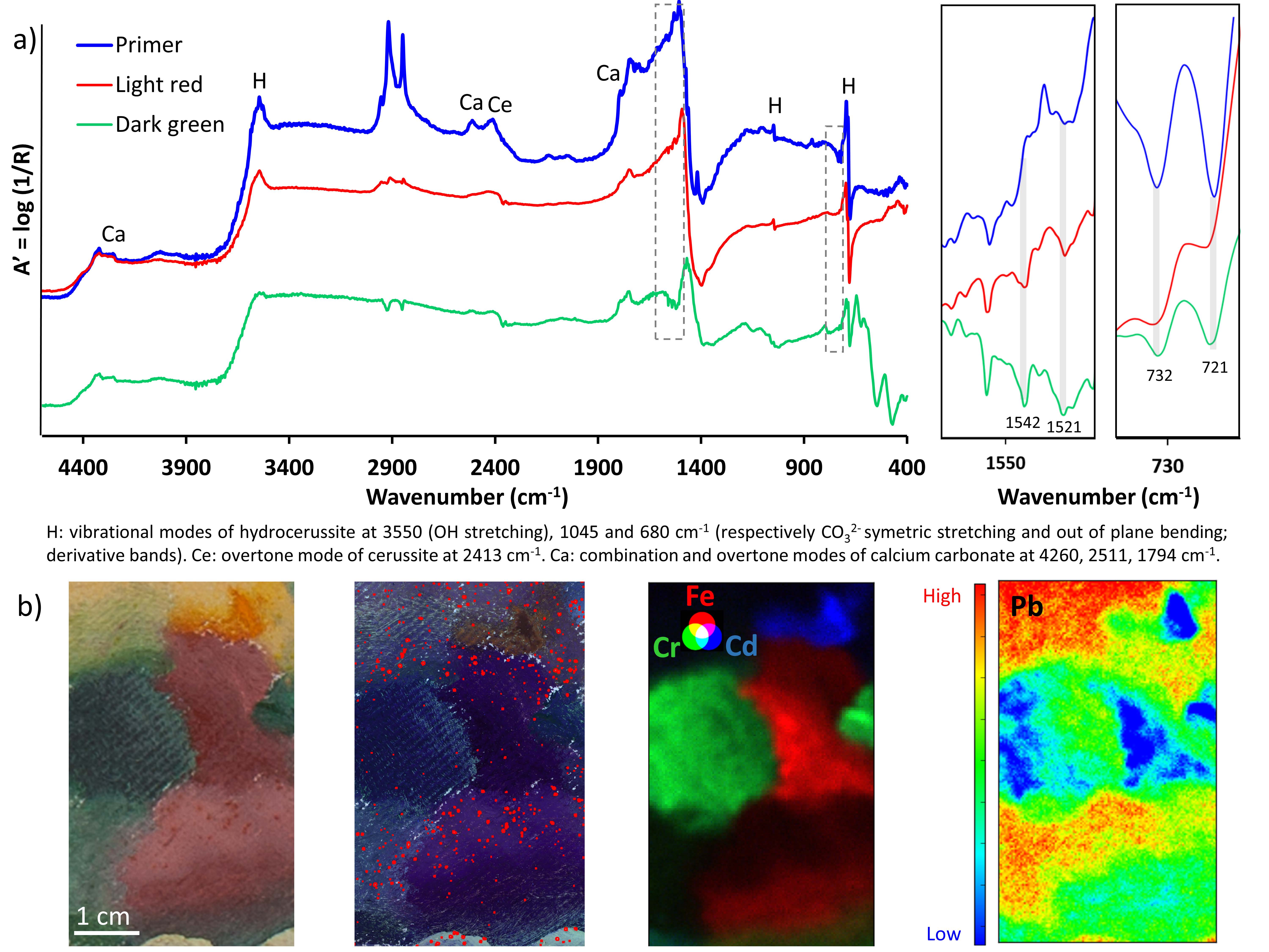}
  \caption{Correlation between the distribution of protusions and of painting materials using complementary non-invasive analytical techniques. a) FTIR spectra in reflectance of the primer (in blue), of the light red area (in red) and of the dark green area (in green). Detail of the $1600-1400^{-1}$ region of the metal carboxylates bands; b) XRF mapping of the studied area (from left to right): visible image of the area; UV-albedo image of the area with overlays of the protrusions in yellow; false color elemental map distribution of chromium (in green), iron (in red) and cadmium (in blue); temperature elemental map distribution of zinc and lead -  in the two later maps, red indicates highest concentration and blue lowest.}
\end{figure}

In all areas analyzed by FTIR, hydrocerussite and metal soap carboxylates were identified (figure 5a). Furthermore, the asymmetric carboxylate bands exhibit similar spectral features for the different colors suggesting the presence of the same type of metal carboxylates (Robinet and Corbeil 2003; Hermans et al. 2014; Hermans et al. 2015) in both the protrusion-rich and protrusion-poor regions. These results highlight that while metal soaps are ubiquitous throughout the paint, they do not always aggregate into protrusions. Instead, paint layers with specific pigment compositions play a decisive role in whether or not surface protrusions form. 

To investigate if protrusion formation is correlated with the presence of certain pigments or paints, the elemental composition was recorded via XRF mapping in the same region as the PS measurements. Elemental distributions of iron, chromium and cadmium may be seen in figure 5b. These data indicate that red iron oxide pigments were employed for the red areas, chromium-based green pigments for the greens and cadmium-based pigments for the yellow and orange. The distribution map of lead indicates that lead white was likely mixed with the colored pigments in varying proportions to produce lighter tonalities. The regions that are richest in lead correlate well with the areas with a higher occurrence of surface protrusions. These observations suggest that the specific type of lead white paint formulation in tubes employed by O’Keeffe and its proportion within the paint play a critical role in the development of surface protrusions, likely related to the aggregation of lead soaps. The latter would be also consistent with the observed UV-induced fluorescence response of the protrusions as reported for lead soap (Higgitt et al. 2003; Keune 2005; Keune and Boon 2007). It is currently being investigated whether the formulation of lead white tube paints O’Keeffe consistently used throughout several decades of her long career might be responsible for the observed abundance of soap protrusions and whether the same correlation in the distribution of lead and surface protrusions is observed on other O’Keeffe paintings.

\section{Conclusions}

Based on the case study of Pedernal, a Georgia O’Keeffe painting exhibiting high number of protrusions, this paper presents an innovative off-the-shelf methodology, based on imaging photometric stereo acquisition under UV illumination, to characterize the 2D-distribution and occurrence of the surface protrusions at the macroscale over an entire artwork. The two detection methods described show that the automated method of analysis produces similar protrusion distribution to the manual method but with decreased precision and substantial underestimation of protrusion numbers when parameters are adjusted to avoid false positives. The developed methodology also demonstrates that the uneven distribution of protrusions can be correlated to the paint composition: in the case study of Pedernal, the findings suggests that the development of protrusions, more numerous in the lead-rich areas, is primarily related to the lead white paint tube formulation in the paint layers, rather than the commercially primed canvas (containing mostly calcium and barium compounds, with only minor amounts of lead), which was previously thought to be the culprit. As an outlook, future research will focus on developing a more robust algorithm foor automated protrusion counting and on providing user-friendly platforms so that the described imaging methodology can be widely applied to following the dynamic evolution of protrusions and thus directly interrogate the myriad factors that may lead to their development.

\acknowledgement

Research at the NU-ACCESS is supported by generous grant from the Andrew W. Mellon Foundation. We thank Dr. Emeline Pouyet, NU-ACCESS, for her assistance with processing the XRF data; Dr. Francesca Casadio, Art Institute of Chicago, for discussing the interpretation of FTIR spectra; and Dr. Xiang Huang, Argonne National Laboratory, for productive discussions.

\bibliography{}
\setlength{\parindent}{0em}
\setlength{\parskip}{1em}
\renewcommand{\baselinestretch}{0}
Bradski G. (2000) The OpenCV library. Doctor Dobb's Journal 25 (11):120-126. http://opencv.org/

Centeno S., Mahon D. (2009) The chemistry of aging in oil paintings: metal soaps and visual changes. The Metropolitan Museum of Art Bulletin 67 (1):12-19

Cossairt O., Huang X., Matsuda N., Stratis H., Broadway M., Tumblin J., Bearman G., Doehne E., Katsaggelos A., Walton M. Surface shape studies of the art of Paul Gauguin. In: Guidi G, Scopigno R, Torres JC et al. (eds) International Congress on Digital Heritage Granada, 28 September-2 October 2015 2015. IEEE, pp 13-20

Cotte M., Checroun E., Susini J., Walter P. (2007) Micro-analytical study of interactions between oil and lead compounds in paintings. Appl Phys A 89 (4):841-848

Duffy M., Martins A., Boon J. (2014) Metal soaps and visual changes in a painting by René Magritte - The Menaced Assassin, 1927. In: van den Berg K.J., Burnstock A., de Keijzer M. et al. (eds) Issues in Contemporary Oil Paint. Springer, Dordrecht, pp 197-203

Duffy S., Bryan P., Earl G., Beale G., Pagi H., Kotoula E. (2013) Multi-light imaging for heritage applications. English Heritage, London

Earl G, Basford P, Bischoff A, Bowman A, Crowther C, Dahl J, Hodgson M, Isaksen L, Kotoula E, Martinez K, Pagi H, Piquette K Reflectance transformation imaging systems for ancient documentary artefacts. In: Dunn S, Bowen J, Ng K (eds) Electronic Visualisation and the Arts (EVA 2011), London, 6–8 July, 2011. BCS, The Chartered Institute for IT, pp 147–154

Earl G, Beale G, Martinez K, Pagi H Polynomial texture mapping and related imaging technologies for the recording, analysis and presentation of archaeological materials. In: Mills J, Barber D, Miller P, Newton I (eds) Proceedings of the ISPRS Commission V Midterm Symposium, Newcastle upon Tyne, 21-24 June, 2010a. pp 218-223

Earl G, Martinez K, Malzbender T (2010b) Archaeological applications of polynomial texture mapping: analysis, conservation and representation. Journal of Archaeological Science 37 (8):2040-2050

Fattal R, Agrawala M, Rusinkiewicz S (2007) Multiscale shape and detail enhancement from multi-light image collections. ACM Transactions on Graphics 26 (3):Paper 51

Faubel W, Simon R, Heissler S, Friedrich F, Weidler PG, Becker H, Schmidt W (2011) Protrusions in a painting by Max Beckmann examined with confocal μ-XRF. Journal of Analytical Atomic Spectrometry 26 (5):942-948

Ferreira E, Boon J, Stampanoni M, Marone F Study of the mechanism of formation of calcium soaps in an early 20th-century easel painting with correlative 2D and 3D microscopy. In: Bridgland J (ed) ICOM Committee for Conservation 16th Triennial Meeting, Lisbon, 19-23 September, 2011. Paper 1604. 

Ferreira E, Gros D, Wyss K, Scherrer N, Zumbühl S, Marone F (2015) Faded shine…. The degradation of brass powder in two nineteenth century paintings. Heritage Science 3:24

Frankot RT, Chellappa R (1988) A method for enforcing integrability in shape from shading algorithms. IEEE Transactions on pattern analysis and machine intelligence 10 (4):439-451

Glassner AS (1995) Chapter 17 The Radiance equation. In: Barsky B (ed) Principles of digital image synthesis, vol 2. Morgan Kaufmann, San Francisco, pp 871-882

Gordon H, Voss K, Kilpatrick K (1993) Angular distribution of fluorescence from phytoplankton. Limnology and oceanography 38 (7):1582-1586

Heeren M, Boon J, Noble P, Wadum J Integrating imaging FTIR and secondary ion mass spectrometry for the analysis of embedded paint cross-sections. In: Vontobel R (ed) ICOM Committee for Conservation 12th Triennial Meeting, Lyon, 29 August-3 September, 1999. pp 228-233

Helwig K, Poulin J, Corbeil M-C, Moffatt E, Duguay D (2014) Conservation issues in several twentieth-century canadian oil paintings: the role of zinc carboxylate reaction products. In: van den Berg KJ, Burnstock A, de Keijzer M et al. (eds) Issues in contemporary oil paint. Springer, Dordrecht, pp 167-184

Hermans J, Keune K, van Loon A, Stols-Witlox M, Corkery, R, and Iedema P The synthesis of new types of lead and zinc soaps: A source of information for the study of oil paint degradation. In Bridgland J (ed), ICOM Committee for Conservation 17th Triennial Conference Preprints, Melbourne, 15-19 Septembe,  2014. pp1604-1612

Hermans J, Keune K, van Loon A, Iedema P (2015) An infrared spectroscopic study of the nature of zinc carboxylates in oil paintings. Journal of analytical atomic spectrometry 30(7):1600-1608

Higgitt C, Spring M, Saunders D (2003) Pigment-medium interactions in oil paint films containing red lead or lead-tin yellow. National Gallery Technical Bulletin 24:75-95

Huang X, Walton M, Bearman G, Cossairt O Near Light Correction for Image Relighting and 3D Shape Recovery. In: Guidi G, Scopigno R, Torres JC et al. (eds) International Congress on Digital Heritage Granada, 28 September-2 October 2015, 2015. IEEE, pp 215-222

Jones R, Townsend J, Stonor K, Duff N Lead soap aggregates in sixteenth and seventeeth century British paintings. In: Parkin HM (ed) AIC Annual Meeting 2006, Paintings Specialty Group Postprints, Providence, Rhode Island, 16-19 June, 2007. pp 24–32

Keck C (1947) Letter from Caroline Keck to Georgia O’Keeffe,  June 17, 1947,  Georgia O’Keeffe Museum Research Center. Copy, gift of the Georgia O’Keeffe Foundation.

Keune K (2005) Binding medium, pigments and metal soaps characterised and localised in paint cross-sections. PhD thesis, University of Amsterdam

Keune K, Boon J (2007) Analytical imaging studies of cross-sections of paintings affected by lead soap aggregate formation. Studies in conservation 52 (3):161-176

Keune K, Ferreira E, Boon J Characterisation and localization of the oil-binding medium in paint cross-sections using imaging secondary ion mass spectrometry. In: Verger I (ed) ICOM Committee for Conservation 14th Triennial Meeting, The Hague, 12-16 September, 2005. pp 796-802

Kratohvil J, Lee M-P, Kerker M (1978) Angular distribution of fluorescence from small particles. Applied optics 17 (13):1980-1982

Ma W-C, Hawkins T, Peers P, Chabert C-F, Weiss M, Debevec P Rapid acquisition of specular and diffuse normal maps from polarized spherical gradient illumination. In: Kautz J, Pattanaik S (eds) Proceedings of the 18th Eurographics conference on Rendering Techniques, Grenoble, 25-27 June, 2007. Eurographics Association, pp 183-194

Maines C, Rogala D, Lake S, Mecklenburg M (2011) Deterioration in abstract expressionistic paintings: analysis of zinc oxide paint layers in works from the collection of the Hirshhorn Museum and Sculpture Garden, Smithsonian Institution. Materials Research Society Symposium Proceedings 1319:275-284

Malesevic B, Obradovic R, Banjac B, Jovovic I, Makragic M (2013) Application of polynomial texture mapping in process of digitalization of cultural heritage. arXiv preprint arXiv:13126935

Malzbender T, Gelb D, Wolters HJ Polynomial texture maps. In: Pocock L (ed) Proceedings of the 28th annual Conference on Computer Graphics and Interactive Techniques (SIGGRAPH 2001), Los Angeles, 12-17 August, 2001. ACM, pp 519-528
Miliani C, Rosi F, Daveri A, Brunetti BG (2012) Reflection infrared spectroscopy for the non-invasive in situ study of artists' pigments. Appl Phys A 106:295-307

Mudge M, Malzbender T, Chalmers A, Scopigno R, Davis J, Wang O, Gunawardane P, Ashley M, Doerr M, Proenca A, Barbosa J (2008) Image-based empirical information acquisition, scientific reliability, and long-term digital preservation for the natural sciences and cultural heritage. Eurographics Tutorials 2008, 29th annual conference of the European Association for Computer Graphics, Crete, 14-18 April. 

Mudge M, Malzbender T, Schroer C, Lum M New Reflection Transformation Imaging Methods for Rock Art and Multiple-Viewpoint Display. In: IEEE Conference on Visual Analytics Science and Technology (IEEE VAST 2006), Baltimore, 31 October-2 November, 2006. pp 195-202

Mudge M, Voutaz J-P, Schroer C, Lum M Reflection Transformation Imaging and Virtual Representations of Coins from the Hospice of the Grand St. Bernard. In: Mudge M, Ryan N, Scopigno R (eds) VAST'05 Proceedings of the 6th International conference on Virtual Reality, Archaeology and Intelligent Cultural Heritage Pisa, 8-11 November, 2005. Eurographics Association, pp 29-40

Noble P, Boon J Metal soap degradation of oil paintings: aggregates, increased transparency and efflorescence. In: Parkin HM (ed) AIC Annual Meeting 2006, Paintings Specialty Group Postprints, Providence, Rhode Island, 16-19 June, 2007. pp 5-19
Noble P, Boon J, Wadum J (2002) Dissolution, aggregation and protrusion. Lead soap formation in 17th century grounds and paint layers. ArtMatters 1:46-61

Noble P, van Loon A, Boon J Chemical changes in Old Master paintings II: darkening due to increased transparency as a result of metal soap formation. In: Verger I (ed) ICOM Committee for Conservation 14th Triennial Meeting, The Hague, 12-16 September, 2005. pp 496-503

O'Donoghue E, Johnson A, Mazurek J, Preusser F, Schilling M, Walton M (2006) Dictated by media: conservation and technical analysis of a 1938 Joan Miró canvas painting. Studies in Conservation 51 (sup2):62-68

Osmond G (2014) Zinc white and the influence of paint composition for stability in oil based media. In: van den Berg KJ, Burnstock A, de Keijzer M et al. (eds) Issues in contemporary oil paint. Springer, Dordrecht, pp 263-281

Osmond G, Boon J, Puskar L, Drennan J (2012) Metal stearate distributions in modern artists' oil paints: surface and cross-sectional investigation of reference paint films using conventional and synchrotron infrared microspectroscopy. Applied Spectroscopy 66 (10):1136-1144

Osmond G, Keune K, Boon J (2005) A study of zinc soap aggregates in a late 19th century painting by R.G. Rivers at the Queensland Art Gallery. AICCM Bulletin 29 (1):37-46

Padfield J, Saunders D, Malzbender T Polynomial texture mapping: a new tool for examining the surface of paintings. In: Verger I (ed) ICOM Committee for Conservation 14th Triennial Meeting, The Hague, 12-16 September, 2005. pp 504-510

Plater M, De Silva B, Gelbrich T, Hursthouse M, Higgitt C, Saunders D (2003) The characterisation of lead fatty acid soaps in ‘protrusions’ in aged traditional oil paint. Polyhedron 22 (24):3171-3179

Robinet L, Corbeil M-C (2003) The characterization of metal soaps. Studies in Conservation 48:23-40

Rogala D, Lake S, Maines C, Mecklenburg M (2010) Condition problems related to zinc oxide underlayers: examination of selected abstract expressionist paintings from the collection of the Hirshhorn Museum and Sculpture Garden, Smithsonian Institution. Journal of the American Institute for Conservation 49:96-113

Rost FW (1995) Fluorescence microscopy, vol 2. Cambridge University Press, Cambridge

Sato I, Okabe T, Sato Y Bispectral photometric stereo based on fluorescence. In: 2012 IEEE Conference proceedings on Computer Vision and Pattern Recognition (CVPR), Providence, Rhode Island, 16-21 June, 2012. IEEE, pp 270-277

Schindelin J, Rueden CT, Hiner MC, Eliceiri KW (2015) The ImageJ ecosystem: An open platform for biomedical image analysis. Molecular reproduction and development 82 (7-8):518-529

Schneider CA, Rasband WS, Eliceiri KW (2012) NIH Image to ImageJ: 25 years of image analysis. Nature methods 9 (7):671-675
Shimadzu Y, Van den Berg JD (2006) On metal soap related colour and transparency changes in a 19th C painting by Millais. In: Boon J, Ferreira E (eds) Reporting highlights of the De Mayerne programme. Netherlands organization for scientific research (NWO), The Hague, pp 43-52

Solé V, Papillon E, Cotte M, Walter P, Susini J (2007) A multiplatform code for the analysis of energy-dispersive X-ray fluorescence spectra. Spectrochimica Acta Part B: Atomic Spectroscopy 62 (1):63-68

Spring M, Ricci C, Peggie D, Kazarian S (2008) ATR-FTIR imaging for the analysis of organic materials in paint cross sections: case studies on paint samples from the National Gallery, London. Anal Bioanal Chem 392 (1-2):37-45

Treibitz T, Murez Z, Mitchell BG, Kriegman D Shape from fluorescence. In: Fitzgibbon A, Lazebnik S, Perona P, Sato Y, Schmid C (eds) 12th European Conference on Computer Vision (ECCV 2012), Florence,  7-13 October, 2012. Springer, pp 292-306
van der Weerd J (2002) Microspectroscopic analysis of traditional oil paint. PhD thesis, University of Amsterdam

Woodham RJ Photometric stereo: A reflectance map technique for determining surface orientation from image intensity. In: Proceedings of SPIE's 22nd Annual Technical Symposium, 1978. International Society for Optics and Photonics, pp 136-143

\end{document}